\documentstyle[epsf]{elsart}
\newcommand{\eq}{\begin{eqnarray}}
\newcommand{\en}{\end{eqnarray}}
\newcommand{\gsim}{\mathrel{
\rlap{\lower4pt\hbox{\hskip1pt$\sim$}}
\raise1pt\hbox{$>$}}}
\newcommand{\lsim}{\mathrel{\rlap{\lower4pt\hbox{\hskip1pt$\sim$}}
\raise1pt\hbox{$<$}}}

\begin{document}
\begin{frontmatter}
\vspace*{-1cm}
\hfill{\large\bf Preprint QFT-TSU/97-50}\\
\vspace*{1cm}
\title{Perturbative framework for the study of \\the properties
of the $\pi^+\pi^-$ atom}
\author[dubna,tomsk]{V. E. Lyubovitskij},
\author[lvta,tbili]{E. Z. Lipartia}
\author[dubna,tbili]{and A. G. Rusetsky}
\address[dubna]{Bogoliubov Laboratory of Theoretical Physics, \\
Joint Institute for Nuclear Research, 141980 Dubna, Russia}
\address[lvta]{Laboratory for Computational Technique and Automation, \\
Joint Institute for Nuclear Research, 141980 Dubna, Russia}
\address[tomsk]{Department of Physics, Tomsk State University,\\
634050 Tomsk, Russia}
\address[tbili] {IHEP, Tbilisi State University, 380086 Tbilisi, Georgia}
%
\begin{abstract}
The perturbative framework is developed for the calculation of the
$\pi^+\pi^-$ atom characteristics (energy level shift and lifetime)
on the basis of the field-theoretical Bethe-Salpeter approach.
A closed expression for the first-order correction to the $\pi^+\pi^-$
atom lifetime has been obtained.
\end{abstract}
\begin{keyword}
Hadronic atoms; Bethe-Salpeter equation; lifetime; energy level shift\\
{\sc PACS}: 03.65.Ge, 03.65.Pm, 13.75.Lb, 14.40.Aq
\end{keyword}
\end{frontmatter}
%
\baselineskip 20pt

Recently a number of experiments on the study of
hadronic atoms have been carried out. Namely, the first estimate of
the $\pi^+\pi^-$  atom lifetime was given in Ref.~\cite{NEMENOV}.
The measurement of the characteristics of the pionic hydrogen~\cite{SIGG}
and pionic deuterium~\cite{CHATELLARD} have been performed.
At present, the DIRAC col\-la\-bo\-ra\-ti\-on is preparing the experiment at CERN
on the high precision measurement of the lifetime of $\pi^+\pi^-$ atoms.
This experiment might provide a decisive improvement in the direct
determination of the difference of the $S$-wave $\pi\pi$ scattering
lengths $a_0^0-a_0^2$ and thus serve as a valuable test for the
predictions of Chiral Perturbation Theory~\cite{CHIRAL}.
Note that the analogous experiments on the observation of the $\pi K$,
$p K$ atoms are also planned. In the view of these experiments there
arises a need in the theoretical framework which would enable one to
calculate the characteristics of such atoms with a high precision based
on the ideas of standard model.

The history of the theoretical study of hadronic atoms has started
from Ref.~\cite{DESER}. In this paper the relations of the energy level
displacement of the $p \pi$ atom due to strong interactions and its
lifetime with the strong $\pi N$ scattering lengths have been established
in the framework of the nonrelativistic scattering theory. In the
following, this approach has been generalized to the case of the
$\pi^+\pi^-$ atom~\cite{URETSKY,BILENKY}. In particular the expression
for the lifetime $\tau_1$ of the $\pi^+\pi^-$ atom in the ground state
was found to be (see also Ref.~\cite{DESER})
\eq\label{fdeser}
\frac{1}{\tau_{1}}=\frac{16\pi}{9}\sqrt{\frac{2\Delta m_\pi}{m_\pi}}
(a^0_0-a^2_0)^2|\Psi_1(0)|^2
\en
\noindent where the isotopic invariance of strong interactions was assumed.
Here we define
$\Delta m_\pi$ is the $m_{\pi^\pm}-m_{\pi^0}$ mass difference, and
$\Psi_1(0)\equiv \phi_0=(m_\pi^3\alpha^3/8\pi)^{1/2}$
is the Coulombic wave function (w.f.) of the pionium
at the origin.

The approach to the study of the problem of hadronic atoms, developed in
Ref.~\cite{DESER}, makes use of the general characteristic feature of
the hadronic atoms~-- the factorization of strong and electromagnetic
interactions. Namely, since the Bohr radii of the atoms composed
from the light mesons and nucleons is of order of a few hundreds fm,
their energy spectrum is almost completely determined by the static
Coulombic potential acting between the constituents. On the other hand,
the decays of hadronic atoms are governed by the strong interactions
which, e.g. for the case of the pionium are responsible for the transition
of the $\pi^+\pi^-$-pair into $\pi^0\pi^0$.
In the following we shall
consider this particular case.
The formula (\ref{fdeser})
demonstrates this factorization property explicitly, expressing the atom
lifetime as a product of two factors~-- the Coulombic w.f. at the
origin and the strong interaction part, completely con\-cen\-tra\-ted in
the $\pi\pi$ strong scattering lengths.

The problem of evaluation of the electromagnetic and strong corrections
to the basic formula (\ref{fdeser}) within different approaches has
been addressed in Refs.~\cite{TRUEMAN}-\cite{KURAEV}. For a brief review
see Ref.~\cite{LUBOVIT}. In this paper within the Bethe-Salpeter (BS)
approach we have derived the relativistic analogue of the formula
(\ref{fdeser}) taking into account the strong interaction corrections
in the first order. These corrections were found to be of the relative
order $10^{-3}$. It should be stressed that the field-theoretical
approaches \cite{SILAGADZE,LUBOVIT,KURAEV} to the problem, unlike the
potential treatment~\cite{TRUEMAN,RASCHE}, do not refer to the concept
of the phenomenological strong interaction $\pi\pi$ potential, which is
a source of an additional ambiguity in the calculations of hadronic atom
characteristics. In the former approaches these characteristics are
expressed directly in terms of the underlying strong interaction (chiral)
Lagrangian, and the results can be compared to the ex\-pe\-ri\-ment, providing
the consistent test of the predictions of chiral theory.

In the present work we suggest a relativistic perturbative framework
for the calculation of the bound-state characteristics of hadronic
atoms (energy levels and lifetime). Our framework, based on the
BS approach to the bound-state problem, is quite similar to the one
used in the treatment of the positronium problem (see, e.g~\cite{LEPAGE}),
and most of the methods which are used in the latter case can be applied
to the hadronic atoms as well. The main purpose of this work is to
demonstrate the possibility (not only in the potential scattering theory,
but in the BS treatment as well) of the clear-cut factorization of strong
and electromagnetic interactions in the observable characteristics of
hadronic atoms, avoiding the double-counting problem in the calculation
of these quantities. One should note that the suggested approach
allows to cal\-cu\-la\-te strong and electromagnetic corrections in all orders
of the perturbation theory. At the present stage we apply the general
formalism to the calculation of the first-order strong and electromagnetic
corrections to the pionium li\-fe\-ti\-me. The results for strong corrections
obtained in Ref.~\cite{LUBOVIT} are reproduced in these calculations.

\vspace*{2mm}
Now we pass to the description of the per\-tur\-ba\-tive framework.
The per\-tur\-ba\-tion expansion is performed around the solution
of the BS equation with the purely
Coulombic kernel similar to that introduced in Ref.~\cite{BARBIERI}
\eq\label{vc}
V_C({\bf p},{\bf q})=\sqrt{w({\bf p})}
\frac{4im_\pi e^2}{({\bf p}-{\bf q})^2}\sqrt{w({\bf q})}
\en
\noindent Here $m_\pi$ denotes the mass of the charged $\pi$-meson, and
$w({\bf p})=\sqrt{m^2_\pi+{\bf p}^2}$.
The factor $\sqrt{w({\bf p})w({\bf q})}$ introduced in the definition
of the instantaneous Cou\-lombic kernel (\ref{vc}) enables one to reduce
the BS equation with such kernel to the exactly solvable Schr\"odinger
equation with the Coulombic potential.
Then, the exact solution of the BS equation
with this kernel is written in the form
\eq\label{wfc}
\psi_C(p)=iG_0(M^\star;p)\,4\sqrt{w({\bf p})}\,\,
\frac{4\pi\alpha m_\pi\phi_0}{{\bf p}^2+\mu^2}\, , \,\,\,\,\,
\bar\psi_C(p)=\psi_C(p)
\en
\noindent where $\mu=m_\pi\alpha/2$ and ${M^\star}^2=m_\pi^2(4-\alpha^2)$
is the eigenvalue corresponding to the unperturbed ground-state solution.
$G_0$ denotes the free Green's function of the $\pi^+\pi^-$-pair.
The BS w.f. (\ref{wfc}) is normalized in the usual way~\cite{NAKANISHI}
\eq\label{norm}
<\psi_C|N(M^\star)|\psi_C>=1\, , \,\,\,\,
N(M^\star)=\frac{i}{2M^\star}\frac{\partial}{\partial M^\star}G_0(M^\star)
\en
The exact Green's function corresponding to the Coulombic kernel (\ref{vc})
can be constructed according to Ref.~\cite{SCHWINGER}
\eq\label{gc4}
\hspace*{-.7cm}G_C(P^\star;p,q)&=&(2\pi)^4\delta^{(4)}(p-q)G_0(P^\star;p)+
G_0(P^\star;p)T_C(E^\star;{\bf p},{\bf q})
G_0(P^\star;q)
\en
\noindent where $T_C$ is given by
\eq\label{gc3}
T_C(E^\star;{\bf p},{\bf q}) &=& 16 i \pi m_\pi \alpha
\sqrt{w({\bf p})w({\bf q})}\,\,\,\biggl[\frac{1}{({\bf p}-{\bf q})^2}
+ \int_0^1\frac{\nu d\rho \rho^{-\nu}}{D(\rho;{\bf p}, {\bf q})}\biggr]\\
D(\rho;{\bf p}, {\bf q}) &=&
({\bf p}-{\bf q})^2\rho-\frac{m_\pi}{4E^\star}
\biggl(E^\star-\frac{{\bf p}^2}{m_\pi}\biggr)
\biggl(E^\star-\frac{{\bf q}^2}{m_\pi}\biggr)(1-\rho)^2
\nonumber
\en
\noindent where $\nu=\alpha\sqrt{m_\pi/(-4E^\star)}$ and
$E^\star=({P^\star}^2-4m_\pi^2)/(4m_\pi)$.

The full BS equation for the $\pi^+\pi^-$ atom w.f. $\chi(p)$ is
written as
\eq\label{bs}
G_0^{-1}(P;p)\chi(p)=\int\frac{d^4k}{(2\pi)^4}V(P;p,q)\chi(q)
\en
\noindent where $V(P;p,q)$ denotes the full BS kernel which is
constructed from the underlying (effective) Lagrangian according
to the general rules and includes all strong and electromagnetic
two-charged-pion irreducible diagrams. In par\-ti\-cu\-lar, it contains
the diagrams with two neutral pions in the intermediate state which
determine the decay the $\pi^+\pi^-$ atom into $\pi^0\pi^0$. Note
that in addition $V(P;p,q)$ contains the charged pion self-energy
diagrams attached to the outgoing pionic legs (with the relative
momentum $q$), which are two-charged-pion reducible. These diagrams
arise in the definition of the kernel $V(P;p,q)$ because the free
two-particle Green's function is used in the l.h.s. of Eq.(\ref{bs})
instead of the dressed one.
The c.m. momentum squared $P^2$ of the atom has the complex value,
corresponding to the fact that the atom is an unstable system.
According to the conventional parametrization, we can write
$P^2=\bar M^2=M^2-iM\Gamma$ where $M$ denotes the "mass" of the atom,
and $\Gamma$ is the atom decay width.

Further, we introduce the four-point Green's function for the
$\pi^+\pi^- \to \pi^+\pi^-$ transition which, by definition, obeys the
inhomogenuos BS equation
\eq
G(P)=G_0(P)+G_0(P)V(P)G(P)
\en
\noindent This function has a pole in the complex $P^2$ plane at the
bound-state energy. The relation between the exact solution $\chi(p)$
and the Coulombic w.f. $\psi_C$ is given by~\cite{LUBOVIT}
\eq\label{init}
<\chi|=C<\psi_C|\, G_C^{-1}(P^\star)G(P),\,\,\,\,\,
{{P^\star}^2\to{M^\star}^2,~P^2\to\bar M^2}
\en
\noindent where $C$ is the normalization constant.
In what follows we assume that this limiting procedure is performed with
the use of the following prescription~\cite{LUBOVIT}
${P^\star}^2={M^\star}^2+\lambda,\,P^2=\bar M^2+\lambda,\,\lambda\to 0$.
The validity of Eq.(\ref{init}) can be trivially checked, extracting
the bound-state pole in $G(P)$ and using the BS equation for $\psi_C$.
The equation (\ref{init}) then turns into the identity.
\vspace{4mm}

In order to perform the perturbative expansion of the bound-state
cha\-rac\-te\-ris\-tics $M$ and $\Gamma$ around the unperturbed values we,
as in Ref.~\cite{LUBOVIT}, split the full BS kernel $V$ into two parts as
$V=V_C+V'$ and consider $V'$ as a perturbation. Further, we introduce
the projector onto the subspace orthogonal to the ground-space w.f.
$\psi_C$ and the "pole subtracted" Coulombic Green's function
\eq\label{qgr}
Q=1-N(M^\star)|\psi_c><\psi_C|,\,\,\,\,
G_R(P^\star)=G_C(P^\star)-\,i\,
\frac{|\psi_C><\psi_C|}{{P^\star}^2-{M^\star}^2}
\en
It is easy to demonstrate that Eq.(\ref{init}) can be
rewritten in the following form
\eq\label{newwf}
<\chi|=<\chi|N(M^\star)|\psi_C><\psi_C|
\bigl[1+(\Delta G_0^{-1}-V')G_RQ\bigr]^{-1}
\en
\noindent where $\Delta G_0^{-1}=G_0^{-1}(P)-G_0^{-1}(P^\star)$. With
the use of Eq.(\ref{newwf}) the following identity is easily obtained
\eq\label{basic}
<\psi_C|\bigl[1+(\Delta G_0^{-1}-V')G_RQ\bigr]^{-1}
(\Delta G_0^{-1}-V')|\psi_C>=0
\en
\noindent which is an exact relation and serves as a basic equation
for performing the perturbative expansion for the bound-state energy.

The equation (\ref{newwf}) expresses the exact BS w.f. of the atom in
terms of the unperturbed w.f. via the perturbative expansion in the
perturbation potential $V'$. This potential consists of the
following pieces:\\
\hspace*{.5cm}
1. The purely strong part, which is isotopically invariant.
This part survives when the electromagnetic interactions are
"turned off" in the Lagrangian.\\
\hspace*{.5cm}
2. The part, containing the diagrams with the (finite) mass
counterterms which are responsible for the $m_{\pi^\pm}-m_{\pi^0}$
electromagnetic mass difference.\\
\hspace*{.5cm}
3. The part, containing the exchanges of one, two, ... virtual
photons and an arbitrary number of strong interaction vertices.

Note that the terms 1 and 2 are more important due to the following
reasons. The first term includes strong interactions which are
responsible for the decay of the pionium. The second term makes
this decay kinematically allowed due to finite difference of
charged and neutral pion masses. Consequently, it seems to be natural
to consider together the pieces 1 and 2. We refer to the corresponding
potential as $V_{12}$.
The $T$-matrix corresponding to the potential $V_{12}$ is defined by
$T_{12}(P)=V_{12}(P)+V_{12}(P)G_0(P)T_{12}(P)$.
The rest of the potential $V'$ is referred as $V_{3}=V'-V_{12}$.
In what follows we restrict ourselves to the first order in the fine
structure constant $\alpha$, i.e. consider the diagrams with only one
virtual photon contained in $V_3$.

The "regular" part of the Coulombic Green's function (\ref{qgr}) can be
split into two pieces according to $G_RQ=G_0(M^\star)+\delta G$. Here
function $\delta G$ corresponds to the ladder of the exchanged Coulombic
photons and thereby contains explicit powers of $\alpha$. It is given by
the following expression
\eq\label{deltag}
\hspace*{-.7cm}\delta G &=&
i\sqrt{w({\bf p})w({\bf q})}\biggl[\Phi({\bf p}, {\bf q})
- S({\bf p})S({\bf q})
\frac{8}{M^\star}
\frac{\partial}{\partial M^\star}\biggr]
G_0(M^\star,p)G_0(M^\star,q)\nonumber\\
\hspace*{-.7cm}\Phi({\bf p}, {\bf q}) &=& 16\pi m_\pi\alpha
\biggl[\frac{1}{({\bf p} - {\bf q})^2} + I_R({\bf p}, {\bf q})\biggr]
+ (m_\pi\alpha)^{-2} S({\bf p})S({\bf q}) R({\bf p}, {\bf q})
\en
\noindent where $S({\bf p}) = 4\pi m_\pi\alpha\phi_0({\bf p}^2 + \mu^2)^{-1}$,
$R({\bf p}, {\bf q}) = 25 - \sqrt{{8}/{\pi m_\pi\alpha}}
[S({\bf p}) + S({\bf q})]$.
The integral $I_R({\bf p}, {\bf q})$ is given by
\eq
I_R({\bf p}, {\bf q}) = \int\limits^1_0 \frac{d\rho}{\rho}
\,\,\,[D^{-1}(\rho;{\bf p}, {\bf q}) - D^{-1}(0;{\bf p}, {\bf q})],\,\,\,
E^\star=-\frac{1}{4}m_\pi\alpha^2
\en
Returning to the basic equation (\ref{basic}) we expand it in the
perturbative series considering $V_3$ and $\delta G$ as perturbations.
Meanwhile we expand $\Delta G_0^{-1}$ in the Taylor series in
$\delta M=\bar M-M^\star$ and substitute
$\bar M=M^\star+\Delta E^{(1)}+\Delta E^{(2)}-{i}/{2}\,\Gamma^{(1)}-{i}/{2}
\,\Gamma^{(2)}+{(8M^\star)}^{-1}{{\Gamma^{(1)}}^2}+\cdots$. Restricting
ourselves to the first order of the perturbative expansion we arrive at
the following relation
\eq\label{master}
\hspace*{-.7cm}
0&=&-\,2iM^\star\delta M\,-
<\psi_C|G_0^{-1}(M^\star)G_0(\bar M)T_{12}|\psi_C>\\[2mm]
\hspace*{-.7cm}
&+&\frac{(\delta M)^2}{2}<\psi_C|G_0^{-1}(M^\star)
[G_0(\bar M)+G_0(\bar M)T_{12}G_0(\bar M)]
\frac{\partial^2G_0^{-1}(M^\star)}{\partial {M^\star}^2}|\psi_C>
\nonumber\\[2mm]
\hspace*{-.7cm}
&-&<\psi_C|(1+T_{12}G_0(M^\star))V_3(1+G_0(M^\star)T_{12})|\psi_C>
\nonumber\\[2mm]
\hspace*{-.7cm}
&-&<\psi_C|G_0^{-1}(M^\star)
[\delta M\frac{\partial G_0(M^\star)}{\partial M^\star}+
G_0(M^\star)T_{12}G_0(M^\star)]G_0^{-1}(M^\star)\delta GT_{12}|\psi_C>
\nonumber
\en
In the lowest order only the first two terms in Eq.(\ref{master}) survive
and in the second term $G_0^{-1}(M^\star)G_0(\bar M)=1$ can be assumed.
Then we obtain
\eq\label{ourdeser}
\Delta E^{(1)}={\rm Re}\left(\frac{i}{2M^\star}
\frac{T_{12}}{m^2}\phi_0^2\right),\,\,\,\,
-\frac{1}{2}\Gamma^{(1)}={\rm Im}\left(\frac{i}{2M^\star}
\frac{T_{12}}{m^2}\phi_0^2\right)
\en
\noindent
If now we use the local approximation for $T_{12}$, assuming that it does
not depend on the relative momenta, we arrive at the well-known Deser-type
formulae for the energy-level displacement and lifetime~\cite{DESER}.
Note that on the mass shell
\eq
\hspace*{-.7cm}
{\rm Re}(iT_{12})\sim T(\pi^+\pi^-\to\pi^+\pi^-),\,\,\,
{\rm Im}(iT_{12})\sim \sqrt{\Delta m_\pi}|T(\pi^+\pi^-\to\pi^0\pi^0)|^2
\en
At the next step we assume that $V_3=\delta G=0$ and evaluate the remaining
integrals in Eq.(\ref{master}), we arrive at the following result
\eq\label{strong}
\frac{\Gamma^{(2)}}{\Gamma^{(1)}}=
-\frac{9}{8}\frac{\Delta E^{(1)}}{E_1}-0.763\alpha,
\hspace*{1.5cm}E_1=-\frac{1}{4}m\alpha^2
\en
The first term of this expression called "strong correction"
was obtained in our previous paper~\cite{LUBOVIT}. However
in difference with the present derivation in Ref.~\cite{LUBOVIT}
we have used the Born approximation for the calculation of
$\Delta E^{(1)}$, i.e. in Eq.(\ref{ourdeser}) $T_{12}$ was
substituted by $V_{12}$. The last term comes from the relativistic
normalization factor $\sqrt{w({\bf p})w({\bf q})}$ in the
instantaneous Coulombic potential~(\ref{vc}). It arises since
in the local approximation for the amplitude $T_{12}$ the atom decay
width is proportional to the quantity $|\int d^4p/(2\pi)^4\psi_C(p)|^2$.
For the particular choice of the potential (\ref{vc}) it is equal to
$\phi_0^2(1-0.381\alpha)^2/m_\pi$.
Since this correction comes from the Coulombic w.f. of the atom, it
does not depend on the parameters of the strong $\pi\pi$ interaction,
and for this reason it was neglected in Ref.~\cite{LUBOVIT}.
Thus the name "strong" in Ref.~\cite{LUBOVIT} refers to the first-order
corrections which survive in the limit when all terms containing
explicit factor $\alpha$ ($V_3$ and $\delta G$) in the equation
(\ref{master}) as well as electromagnetic (relativistic) correction
coming from the Coulombic w.f., are assumed to vanish.

The last term in Eq.(\ref{master}), proportional to $\delta G$,
corresponds to the correction due to the exchange of the (infinite number)
of Coulombic photons. With the use of the explicit expression for
$\delta G$~(\ref{deltag}) the integrals in this term can be easily
evaluated, using again the local approximation for $T_{12}$. These integrals
are ultraviolet convergent, containing, however, an infrared enhancement
$\alpha{\rm ln}\alpha$ which stems from the infrared-singular one-photon
exchange piece in Eq.(\ref{gc3}). Below we present the result of the
calculations~\footnote{
This correction has been recently calculated by H.Jallouli and
H.Sazdjian in the framework of the three-dimensional constraint
equations (H.Sazdjian, private communication)}:
\eq\label{exchange}
M^\star\delta M\frac{T_{12}}{m^2}\phi_0^2\frac{1}{8E_1}
-i\alpha(-2.694+{\rm ln}\alpha)\frac{1}{16\pi m}T_{12}^2\phi_0^2
\en
Collecting all terms together and using Eqs.(\ref{ourdeser}) for
relating ${\rm Im}T_{12}$ to $\Delta E^{(1)}$, we finally arrive
at the following expression for the first-order correction to the
$\pi^+\pi^-$ atom decay width
\eq\label{final}
\hspace*{-.7cm}\frac{\Gamma^{(2)}}{\Gamma^{(1)}}&=&
\underbrace{-\frac{9}{8}\,\frac{\Delta E^{(1)}}{E_1}}_{\rm strong}\,\,\,
+\underbrace{(-0.763\alpha)}_{\rm kernel~normalization}\, + \,\,\,\,
\underbrace{\left({1}/{2}+2.694-{\rm ln}\alpha\right)
\frac{\Delta E^{(1)}}{E_1}}_{\rm Coulombic~photon~exchanges}\,\,\,
-\nonumber\\
\hspace*{-.7cm}&-&{\left( M^\star\Gamma^{(1)}\right)^{-1}}
{\rm Re}<\psi_C|(1+T_{12}G_0(M^\star))V_3(1+G_0(M^\star)T_{12})|\psi_C>
\en
\noindent
Let us now turn to the discussion of the last term in Eq.(\ref{final}).
The potential $V_3$ present in this term contains the diagrams of two
types: a) the diagrams where $\pi^+$ and $\pi^-$ lines are connected by
photons, b) the diagrams where these lines are connected by strongly
interacting particles ($\pi$, $\rho$, ...) as well. The diagrams of the
first type correspond to the retardation correction \cite{SILAGADZE},
correction due to vacuum polarization \cite{EFIMOV}, etc. The diagrams of
the second type correspond to the radiative corrections \cite{KURAEV}.
In the Eq.(\ref{final}) all these corrections are given in a closed form
avoiding any difficulties connected with double counting problem. The
kernel $(1+T_{12}G_0(M^\star))V_3(1+G_0(M^\star)T_{12})$ which appears
in this term is constructed from the underlying Lagrangian with the use
of the conventional Feynman diagrammatic technique. Note that the matrix
element
\eq
<\psi_C|(1+T_{12}G_0(M^\star))V_3(1+G_0(M^\star)T_{12})|\psi_C>
\en
can be written as $<\psi_{12}|V_3|\psi_{12}>$ 
where $\psi_{12} = [1+G_0(M^\star)T_{12}] \psi_C$ stands for the solution
of the BS equation with the kernel $\Delta V = V - V_3$. The detailed
analysis of the above mentioned corrections will be addressed in our future
publications.

Apart of these corrections there are two types of corrections implicit
in for\-mu\-lae (\ref{ourdeser}). The first is due to the finite
difference of the masses of charged and neutral pions. In order to
evaluate this correction one has to calculate the real and the imaginary
parts of the $\pi^+\pi^-\to\pi^+\pi^-$ transition matrix $T_{12}$ at the
two-charged-pion threshold with the use of the low-energy chiral Lagrangian
\cite{CHIRAL}, in two cases: with and without the term responsible
for the $m_{\pi^\pm}-m_\pi^0$ mass splitting. The second correction is
caused by the dependence of the transition matrix $T_{12}$ on the
relative momenta of $\pi^+\pi^-$ pair. This correction can be
evaluated with the use of Eq.(\ref{ourdeser}) provided the explicit
expression of $T_{12}$ is known.

In order to estimate the size of the first three terms in
Eq.(\ref{final}) we have used the following value of the singlet
scattering length $m_\pi(2a_0^0+a_0^2)=0.49$ \cite{DUMBRAJS} corresponding
to the value $\Delta E^{(1)}/E_1=0.24 \%$. The first, second and third terms
then contribute, respectively, $-0.26\%$, $-0.55\%$ and $+1.85\%$, to
the decay width, and their total contribution amounts up to $\sim 1\%$.
The largest contribution comes from the $\alpha{\rm ln}\alpha$ term
in Eq.(\ref{final}). Note that this logarithmic term is likely to
cancel with the similar term coming from the last piece in
Eq.(\ref{final}) in analogy with the positronium case \cite{LEPAGE}.

We thank G.V. Efimov, J. Gasser, M.A. Ivanov, E.A. Kuraev, H. Leutwyler,
P. Minkowski, L.L. Nemenov, H. Sazdjian and A.V. Tarasov for useful
discussions, comments and remarks. A.G.R. thanks Bern University for
the hospitality where a part of this work was completed.
This work was supported in part by the INTAS Grant 94-739,
by the Russian Fund of Fundamental Research (RFFR) under contract
96-02-17435-a.

\end{document}